\documentclass[twocolumn,prl,nofootinbib]{revtex4}

\usepackage{amsmath,amssymb,amsthm}
\usepackage[dvips]{graphicx}
\usepackage{psfrag}

\newcommand{\Ref}[1]{(\ref{#1})}

\newcommand{\N}{\mathbb{N}}

\newcommand{\R}{\mathbb{R}}
\newcommand{\C}{\mathbb{C}}

\def\be{\begin{equation}}
\def\ee{\end{equation}}
\def\bes{\begin{eqnarray}}
\def\ees{\end{eqnarray}}
\def\arr{\rightarrow}

\def\ra{\rangle}
\def\f{\frac}

\def\what{\widehat}

\newcommand{\SU}{\mathrm{SU}}
\newcommand{\SL}{\mathrm{SL}}

\begin{document}

\title{Quantizing speeds with the cosmological constant}
\author{{\bf Florian Girelli}\footnote{fgirelli@perimeterinstitute.ca},
{\bf Etera R. Livine}\footnote{elivine@perimeterinstitute.ca}}
\affiliation{Perimeter Institute, 35 King Street North, Waterloo, Ontario, Canada N2J 2W9}

\begin{abstract}
Considering the Barrett-Crane spin foam model for quantum gravity with (positive) cosmological constant, we show that
speeds must be quantized and we investigate the physical implications of this effect such as the emergence of an
effective deformed Poincar\'e symmetry.
\end{abstract}

\maketitle



\section{Introduction: Spin Foams at $\Lambda\ne 0$}


One of the most important problem to address in quantum field theory (on curved space-time) and in general relativity
is the cosmological constant ``problem'', and it is usually believed that  quantum gravity should have something to say
about it. Furthermore quantum gravity theories actually often need a non-zero cosmological constant to be well-defined
or in order to have some interesting content. For example the cosmological constant is an essential parameter in
dynamical triangulations and matrix models and gets renormalized. In (canonical) loop quantum gravity, although the
theory is well-defined, the only exact state with semi-classical properties is the so-called Kodama state, which exists
only for $\Lambda\ne 0$ \cite{lee}. Spin foams also get regularized by the cosmological constant, which introduces by
quantum deformation a natural cut-off which makes all amplitudes finite (for reviews see \cite{alej,dan}). One would
also mention doubly special relativity  (DSR) \cite{dsr}, which predicts a effective $\kappa$-deformed Poincar\'e
theory as a flat limit $\Lambda\arr 0$ while keeping $l_P$ finite \cite{flatlim}: the argument needs a consistent
theory defined for $\Lambda\ne 0$. Finally, taking into account the cosmological constant leads naturally to a holographic
bound for the entropy in quantum gravity \cite{holo}. In the present work, we wish to explore some ``physical''
implications of a non-vanishing cosmological constant in the spin foam setting, more precisely within the framework of
the Barrett-Crane model \cite{BC}.

The Barrett-Crane model is a state sum model for $3+1$ quantum gravity.
It is based on the reformulation of general
relativity as a constrained BF theory with gauge group the Lorentz group
$\SL(2,\C)$: BF theory is a topological field
theory which we know how to quantize exactly by discretizing it, then we
impose the constraints leading to general
relativity at the quantum level. As a spin foam model, it defines a discretized
quantum space-time structure.
Interpreting it as defining transition amplitudes between quantum states of
geometry of 3d hypersurfaces/slices, it can
be seen as defining the path integral for a loop-like canonical gravity
theory \cite{covlqg} with quantum states
defined by ``projected'' spin networks \cite{proj} which encode information
about both the Lorentz connection and the
time normal to the hypersurface: the connection is taken into account through
the holonomies along the edges of the
graph underlying the spin network and the time normal defines vectors at the
vertices of the graph valued on the upper
hyperboloid ${\cal H}_+$ of the Minkowski space (timelike future-oriented normalized vectors).

A positive cosmological constant $\Lambda>0$ is then usually taken into account by q-deforming the Lorentz group
$\SL(2,\C)$ into $U_q(\SL(2,\C))$, in a similar way that the Ponzano-Regge model for 3d gravity based on the $\{6j\}$
symbols for $\SU(2)$ becomes the Turaev-Viro model based on $U_q(\SU(2))$. The argument is based on the fact that the
topological BF theory with cosmological constant is known to be quantized using a q-deformation with
$q=\exp(-l_P^2\Lambda)$. Then one can write a quantum Barrett-Crane model \cite{qBC}. Classically, projected spin
networks are defined with $\SL(2,\C)$ (simple unitary) representations on the edges, describing the state of the
Lorentz connection, and ${\cal H}_+$ valued vectors at the vertices. The representations are labelled by a real number
$\rho\ge0$, which describes the area of a surface intersecting the corresponding link of the spin network: ${\cal
A}=l_P^2\rho$. In the q-deformed model, using the representation theory of $U_q(\SL(2,\C))$, it turns out that the
areas become bounded: ${\cal A}\le\pi/\Lambda$. Moreover the time normal
becomes quantized: the hyperboloid ${\cal H}_+$ becomes a pile of fuzzy spheres
with quantized radii. Now, how can we interpret this time normal? It corresponds
to all possible momenta for an object on the hypersurface, thus to all
allowed speeds/velocities: the speed (with
respect to a given observer) is now quantized.
Another effect, which we will not discuss much, is that, as we deal with fuzzy spheres,
components of the speed do not commute, e.g $[v_x,v_y]\ne0$, and thus can not be determined
a priori at the same time.

In the next section, we will give the resulting quantum spectrum of the velocities. And
afterwards, we will discuss possible physical implications of this effect due to
the cosmological constant.

\section{A discrete spectrum of velocities}

The classical hyperboloid ${\cal H}_+$ is the set of vectors such
that $x_0^2-|\vec{x}|^2=1$. The quantum hyperboloid
becomes a stack of fuzzy spheres ($\vec{x}$ becomes a ``fuzzy vector'') with
quantized radii given by:
\be
X_0=\sqrt{\f{1}{2}\left(1+\f{\cosh n\lambda}{\cosh\lambda}\right)},
\label{rule}
\ee
with $n\in\N,n\ge 1$ and $\lambda=\Lambda l_P^2$.

\begin{figure}[t]
\label{hyper}
\begin{center}
\psfrag{x}{$x_0$}
\psfrag{h}{${\cal H}_+$}
\includegraphics[width=5cm]{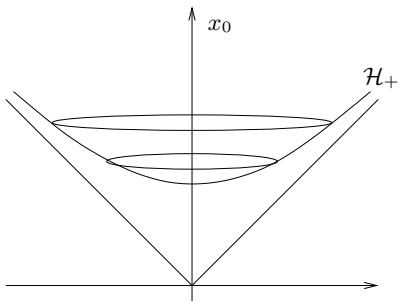}
\end{center}
\caption{Upper hyperboloid ${\cal H}_+$ with quantized levels in $X_0$.}
\end{figure}

Classically, a vector on ${\cal H}_+$ defines a boost or equivalently a boosted observer. Its speed is $v=\tanh\eta$
with the boost parameter defined by $x_0=\cosh\eta$. Resultantly, it seems that speeds in the q-deformed space-time
will be quantized. This effect is not a phenomenon arising from the kinematics of a particle on quantum gravity
background, but directly reflects the discretized structure of space-time. Surely that the inclusion of
particles/fields in spin foams will modify the structure and properties of the partition function, however it is likely
it will not affect the basic structure of the underlying quantum geometry but only modify its detail. From this point
of view, it is natural to expect the discretized time normal structure to be translated into quantized speed for
objects propagating in a medium defined by such a quantum space-time. Then the allowed velocities are labelled by the
integer $n\ge 1$: \be v_n =\sqrt{1-\f{2}{1+\f{\cosh n\lambda}{\cosh\lambda}}} =\sqrt{\f{\cosh
n\lambda-\cosh\lambda}{\cosh n\lambda+\cosh\lambda}}, \ee and depend in a non-trivial fashion on the (reduced)
cosmological constant $\lambda$. The corresponding rapidities are: \be \eta_n=\f{1}{2}\textrm{argch}\left(\f{\cosh
n\lambda}{\cosh\lambda}\right). \label{etan} \ee

The classical limit is recovered by taking $\lambda\arr 0$ while keeping $n\lambda$ finite.
Then when $n\lambda\arr 2l\in\R_+$, $X_0$ goes to $\cosh l$. And we thus reproduce the continuous
hyperboloid.

Given a fixed $\lambda$, the typical behavior is that, for small $n$, $v_n$ grows
slowly, almost linearly, then the
values $v_n$ accumulate near the speed of light $c$ (here set to 1). Getting into
details, when $n\lambda\ll 1$, $v_n$
behaves as: \be v_n\sim\f{\lambda}{2}\sqrt{n^2-1}, \ee that is (almost) linearly when
$n\ge 1$. Let us point out that
the minimal speed ($v_2$) is of order of the reduced cosmological constant $\lambda$,
so that the cosmological constant corresponds somehow to a minimal energy/excitation.

In our physical context, $\lambda$ is very small, in which case the spacing
between consecutive values of speed is very
small and the spectrum looks continuous. If we would like to observe (directly)
this discreteness, we would need to go
to low energies/speeds and not to high velocities (near $c$) where the spectrum
becomes dense. However, in order to see
very low speeds, we would need to measure very small length intervals or distances,
which requires high energy probes.
At intermediate speeds, the normalized spacing $\delta v/v$ will go as $1/n$ and
therefore will decrease very quickly.
Let us point out that although the spacing $\delta v$ between speed levels is almost constant
of order $\lambda$ for small velocities, it is then exponentially decreasing
when approaching the speed of light behaving as $\lambda/\exp(n\lambda)$.

For high values of the cosmological constant, the linear phase gets shorter and
eventually vanishes. When $\lambda\arr
1$ i.e the cosmological length $l_C=1/\sqrt{\Lambda}$ becomes as small as the
Planck length $l_P$, there is just a few
values of the speed before the spectrum saturates at the speed of light $c$: there is
only a very few allowed values of
speed other than $c$!

\begin{figure}[t]
\begin{center}
\includegraphics[height=5cm,width=6.5cm]{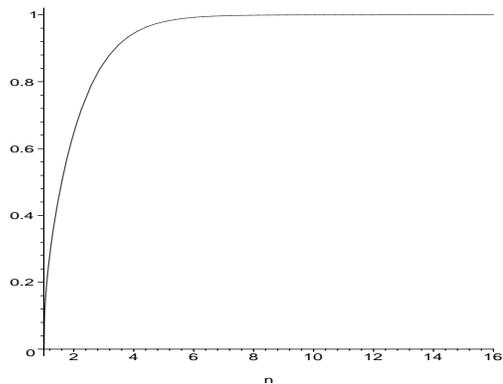}
\end{center}
\caption{Velocity spectrum at $\lambda=1$: it saturates at the speed of light already
at $n\sim 7$.}
\label{lambda1}
\end{figure}

\begin{figure}[t]
\label{lambdap}
\begin{center}
\includegraphics[height=5cm,width=6.5cm]{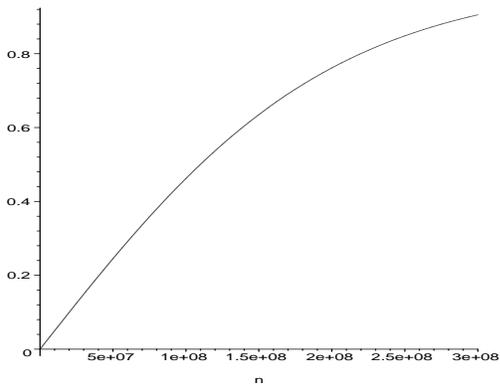}
\end{center}
\caption{Velocity spectrum at $\lambda=10^{-8}\ll 1$: we see the linear regime up
to $n\sim 1/\lambda$, then we reach the bound $c$.}
\end{figure}

\section{Observing quantized speeds}

As mentioned earlier we are not dealing with the kinematics of a particle over  space-time but the deep quantum
structure of space-time itself. Introducing matter in the Barrett-Crane model is still missing and so understand its
dynamics in this context is still lacking. We can however guess that certainly the quantization of the speed  would
enhance some of the effects already predicted, for example in loop quantum gravity \cite{pullin}. We can therefore try
to make some comments on the new aspect of quantum space-time this is shedding light on.

First note that usually the cosmological constant is understood as a global feature
of space-time, it is the vacuum
energy of space-time. This global structure therefore has some very local effects!
It would be interesting to turn this argument around: now the discrete quantum structure of space-time (with
discretized time directions) can be considered as the origin of the cosmological constant. This would provide a new and
original proposition for the nature of the cosmological constant.

We can turn now to the physical consequences of the quantization of speed. Ordinarily the effects of quantum gravity are
expected to occur at very high energy, for example when dealing with Gamma rays bursts \cite{amelino}. In fact until
recently quantum gravity was not expected to be testable due to this high energy problem. If one tries to measure the
quantized speed, then, strangely the effect we want to measure is at very low energy!  The effect that we are
suggesting has therefore the originality to be present at very low energy, contrary to the usual predictions. The
particle must have a very small speed so that we can measure the quantization effect. In fact we would be interested to
measure the difference between the two quanta of speed, and as we have seen that the curve is essentially linear at low
energy, we would want to measure a difference of speed of the order of $\lambda$. It is interesting to try to quantify
the effect that one should expect. $\lambda $ is  of order $10^{-120}$ taking into account the cosmological constant
$\Lambda$ at $10^{-56}\textrm{cm}^{-2}$.

This means that our measurement tool should be of precision of $\lambda$, which is pretty small!  A possible
alternative is to cool down the some particles and then try to measure their position to deduce their speed. Ideally
one would expect to have some resonances which would appear. However the precision of the measurement should be so high
that  the photons would have a huge energy (of order $\lambda^{-1}$). This is not doable either.
From this point of view, it looks as if the spectrum of speeds is simply continuous.

It seems that those effects are not measurable in our days due to the very weak value of the actual cosmological
constant. In order to have some sensible effect, one should consider a bigger cosmological constant and therefore early
cosmology. One can then hope to measure some relics of this effect for example in the cosmological background.

In fact in the very early universe, the quantization of speeds must have played a very important role. Indeed, the
cosmological horizon being very small, the  cosmological constant was then very big and so it is easy to see that very
few speeds were allowed (cf. Fig.\ref{lambda1}) and the difference between them was pretty high. Matter must have then
travelled on some shells of same speed, which might have had some influence on the structure formation. In fact one can
certainly argue that this must have enhanced the ability to condense and to form structures. This would be very
interesting to inquire this effect for example in the WMAP context.


Let us mention another effect of the quantized speeds, which point to a possible link
between the Barrett-Crane model and theories based on non-commutative space-time such
as doubly special relativity. The argument is based on change of observers and the resulting
rule of addition of speeds. For such a purpose, we place ourselves in the case of an effective
theory around a flat spacetime resulting from the spinfoam theory.
Then, let us consider an object C moving with respect to B with rapidity $\eta_n$
and B moving in the same direction with respect to an object A with rapidity $\eta_m$.
We expect from special relativity that the speed of C with respect to A would be
$\eta=\eta_n+\eta_m$. However $\eta$ would need to be also quantized, but considering
the law of quantization of speeds we use, there doesn't
exist any level $p$ such that $\eta=\eta_p$. So it is reasonable to modify the rule of sums
of speeds by summing the levels $n+m$ instead of summing the rapidities $\eta_n+\eta_m$.
An alternative to such modification would be consider the speed
as an operator $\what{v}$ and to allow systems to be in quantum superpositions of
different speeds: then we could respect the usual summing rule of the rapidities allowing
that an object moving at a definite speed $|n\ra$ with respect to an observer would
move in a superposition $\sum_n\alpha_n|n\ra$ with respect to another one. This would mean that
a single object seen by a given observer would seem to be spread for another observer. Such
an effect is expected with wave packets (true quantum objects instead of
idealised pointlike particles), but only for accelerated observers or observers in
a curved spacetime. Nevertheless, in an effective theory of quantum gravity, we migth expect
some macroscopic effects of the quantum fluctuations of the curvature: in this context, a
spreading of a wave packet (a rainbow effect) dependent on the observer sounds plausible.
Still, it seems physically more natural to expect a modification of the rule of composition
of velocities. This seems related to proposed modifications of the action of the Lorentz boosts
and of Poincar\'e translations (possibly to a non-linear action) in the context of doubly
special relativity or more generally for effective theories of quantum gravity using
a non-commutative spacetime.

Finally, let us investigate the implications of the quantization of speeds on
scattering processes. Indeed momenta are now quantized, so we should get some restrictions
on possible physical processes. More precisely, the 4-momentum $p$ is $m(\gamma,\gamma\vec{v})$.
If we consider two particles of the same mass  $m$ with initial momenta
$p_a,p_b$ scattering to final states with momenta $p_k,p_l$ ($a,b,k,l$ being the speed levels
$n$). As we haven't dealt with
the quantization of the direction of the speed $\vec{v}$, let us look at the conservation of
energy: $\gamma_a+\gamma_b=\gamma_k+\gamma_l$. Looking at the quantization rule \Ref{rule} for
$\gamma=\cosh\eta$, it appears that there exists a single solution to this equation given
by $\{a,b\}=\{k,l\}$. Could there exist a modification of this law to allow for more physical
processes? Indeed, when looking at low energy:
$$
n\lambda\ll 1 \quad\Rightarrow\quad \gamma_n-1\approx\f{1}{2}v_n^2\approx\f{1}{8}n^2\lambda^2,
$$ 
so that we can propose a new law of conservation of energy: $a^2+b^2=k^2+l^2$. Let us point out
that now the energy is quantized quadratically (in $n^2$) instead of linearly in $n$.
This implies that the energy spectrum for composite objects will be the sum of the energies
of each component, that is a sum of $n^2$'s, which generates a difference between a fundamental
object ($E\sim n^2$) and a composite one ($E\sim \sum n^2$) at the kinematical level, such as
occuring in doubly special relativity \cite{leejoao}.
This new equation has few solutions for low $a,b$, i.e. few physical processes, but the number
of solutions increase considerably when we get at higher energies, so
that we get a sensible number of possible physical scattering at our scale of energies and
possibly ``new'' interactions at even higher energies.
Such a modification of the law of energy conservation is of the same kind of the modification
of composition of speeds proposed previously.


\section*{Conclusion}


We are proposing one of the first possible experimental predictions of the Barrett-Crane model, that is that speed must
be quantized, and even if a comprehensive model including matter is still lacking, one can argue that those effects
should have been non negligible in the early unverse and then had some consequences on the structure formation. The
natural place to see this effect is of course the cosmological background. This effect is moreover very original as in
our scale it should be visible in the {\it very low energy regime}, instead of the usual very high energy regime.
Furthermore, from a theoretical point of view, it seems that quantized speeds
point towards a link between the quantum
Barrett-Crane model and the effective theories of the kind of doubly special relativity.
Indeed we proposed a modification of the law of composition of speeds and of the law of
conservation of energy, which should be visible at high energies (as long as an effective theory
of quantum gravity for a flat spacetime could be valid at high energies).
A last issue one could keep in mind is
whether all this applies to the case of a negative cosmological constant, which
corresponds to taking a quantum deformation parameter $q=\exp(-\Lambda l_P^2)>1$.

\section*{Acknowledgements}

We would like to thank Giovanni Amelino-Camelia, Daniele Oriti and Lee Smolin for
their interest and their comments.


\begin{thebibliography}{99}

\bibitem{lee}
L Smolin,
{\it Quantum gravity with a positive cosmological constant},
hep-th/020907

\bibitem{alej}
A Perez,
{\it Spin Foam Models for Quantum Gravity},
Class.Quant.Grav. {\bf 20} (2003) R43,
gr-qc/0301113

\bibitem{dan}
D Oriti,
{\it Spacetime geometry from algebra: spin foam models for non-perturbative quantum gravity},
Rept.Prog.Phys. {\bf 64} (2001) 1489-1544,
gr-qc/0106091

\bibitem{dsr}
G Amelino-Camelia,
{\it Relativity in space-times with short-distance structure governed by an observer-independent (Planckian) length scale},
Int.J.Mod.Phys. {\bf D11} (2002) 35-60,
gr-qc/0012051

\bibitem{flatlim}
G Amelino-Camelia, L Smolin, A Starodubtsev,
{\it Quantum symmetry, the cosmological constant and Planck scale phenomenology},
hep-th/0306134

\bibitem{holo}
L Smolin,
{\it A holographic formulation of quantum general relativity},
Phys.Rev. {\bf D61} (2000) 084007,
hep-th/9808191


\bibitem{BC}
JW Barrett, L Crane,
{\it Relativistic spin networks and quantum gravity},
J.Math.Phys. {\bf 39} (1998) 3296-3302,
gr-qc/9709028 \\
JW Barrett, L Crane,
{\it A Lorentzian Signature Model for Quantum General Relativity},
Class.Quant.Grav. {\bf 17} (2000) 3101-3118,
gr-qc/9904025

\bibitem{covlqg}
ER Livine, S Alexandrov,
{\it $\SU(2)$ Loop Quantum Gravity seen from Covariant Theory},
Phys.Rev. {\bf D67} (2003) 044009,
gr-qc/0209105

\bibitem{proj}
ER Livine,
{\it Projected Spin Networks for Lorentz connection: Linking Spin Foams and Loop Gravity},
Class.Quant.Grav. {\bf 19} (2002) 5525-5542,
gr-qc/0207084

\bibitem{qBC}
K Noui, Ph Roche,
{\it Cosmological Deformation of Lorentzian Spin Foam Models},
Class.Quant.Grav. {\bf 20} (2003) 3175-3214,
gr-qc/0211109

\bibitem{pullin}  R Gambini, J Pullin,
 {\it Nonstandard optics from quantum spacetime}, Phys.Rev. D {\bf 59} (1999) 124021,
gr-qc/9809038

\bibitem{amelino} G Amelino-Camelia, {\it Are we at the dawn of quantum-gravity phenomenology?},  Lect.Notes Phys. {\bf 541} (2000) 1-49, gr-qc/9910089

\bibitem{leejoao}
 J Magueijo, L Smolin,
{\it Generalized Lorentz invariance with an invariant energy scale},
Phys.Rev. {\bf D67} (2003) 044017,
gr-qc/0207085



\end{thebibliography}
\end{document}